\documentclass[jounral]{IEEEtran}

\ifCLASSINFOpdf
 \usepackage[pdftex]{graphicx}
 \usepackage{subfloat}
\else
\fi
\usepackage{amsfonts}
\usepackage[cmex10]{amsmath}
\newcommand{\E}{\mathbb {E}}
\newcommand{\R}{\mathbb R}

\renewcommand{\>}{\rangle}

\usepackage[caption=false,font=normalsize,labelfont=sf,textfont=sf]{subfig}

\begin{document}
\title{Simulation of Fractional Brownian Surfaces via Spectral Synthesis on Manifolds}
\author{Zachary~Gelbaum, Mathew~Titus}%
\maketitle
\begin{abstract}
Using the spectral decomposition of the Laplace-Beltrami operator we simulate fractal surfaces as random series of eigenfunctions.  This approach allows us to generate random fields over smooth manifolds of arbitrary dimension, generalizing previous work with fractional Brownian motion with multi-dimensional parameter.  We give examples of surfaces with and without boundary and discuss implementation.
\end{abstract}

%
\IEEEpeerreviewmaketitle
\section{Introduction}
%
%
%
%
\IEEEPARstart{A} common approach to simulating fractal surfaces is via the sample paths of fractional Brownian motions and their multidimensional extensions to $\R^n$ (e.g. \cite{MR952853,franceschetti2006scattering}).  These random fields are self-similar in distribution in that when sampled at various scales the distribution of the sample is the same up to a constant scaling factor.  However, they are indexed by $\R^n$.  Suppose instead one wanted to simulate a fractal surface with some more complex geometry or topology, e.g., a fractal cylinder.  The analogous approach to simulating such an object would require a random field indexed by a manifold, such as a surface in $\R^3$, that possessed the same properties as fractional Brownian motion, i.e., self-similarity, almost surely H\"older continuous sample paths, and stationary increments. 

Constructing such fields has been a subject of ongoing research for a number of years and until recently, extensions were known only for limited classes of surfaces, such as spheres or hyperboloids, and only for certain ranges of self-similarity, i.e., only with Hurst index $\alpha\in(0,1/2]$.  Then in \cite{FBFOM} self-similar Gaussian random fields were constructed over wide classes of surfaces, including arbitrary compact manifolds, both with and without boundary.  These extensions were shown to have H\"older continuous sample paths, be self-similar in distribution, and have stationary increments, thus possessing all the basic properties one would expect of an extension of fractional Brownian motion.  In this paper we describe the simulation of these fields for surfaces in $\R^3$.

The simulation of Gaussian random fields presents non-trivial challenges, in particular if one considers index sets other than $\R^n$.  One common approach is the use of Fourier or wavelet techniques (e.g. \cite{MR952853}, ch. 2, \cite{franceschetti2006scattering}), but if one is working on a general surface the Fourier transform is no longer available.  On the other hand, one can view the Fourier transform as the spectral decomposition of the Laplacian, the sine and cosine functions being the eigenfunctions.  These we do have on a surface, and so the analogous approach is to build and simulate random fields over manifolds and surfaces using Fourier series of eigenfunctions of the Laplace-Beltrami operator.  One novel feature of this approach is that using the Dirichlet Laplacian for a surface with boundary allows us to produce fields with almost sure boundary values.  We begin the next section with a derivation in the simple case of the interval $[0,1]$, before moving on to surfaces.

\section{Fractional Brownian fields over surfaces}
\subsection{A Basic Example}
As motivation and a base case for our construction, consider Brownian motion on $[0,1]$.  Being a Gaussian process, $B_t$ is determined by its covariance function, \[\mathbb E[B_tB_s]= \min\{s,t\}\equiv s\wedge t.\]  Now notice that $s\wedge t$ is the Green's function of the Laplacian on $[0,1]$, $-\Delta=-\frac{d^2}{dx^2}$, i.e., for $f\in L^2[0,1]$ \begin{equation}-\frac{d^2}{ds^2}\int_0^1s\wedge tf(t)dt=f(s).\end{equation}  Of course we must specify boundary conditions to have a well defined Laplacian, and we see that these come from $s\wedge t$: If \[F(s)=\int_0^1s\wedge tf(t)dt\] then $F(0)=0$ and $F^\prime(1)=0$.  Thus we can say that $B_t$ is the unique (up to equality in distribution) Gaussian process on $[0,1]$ whose covariance is the Green's function of $-\Delta$ acting on smooth functions $f:[0,1]\to \R$ such that $f(0)=f^\prime(1)=0$.

If we let $K:L^2[0,1]\to L^2[0,1]$ be given by \[K(f)(s)=\int_0^1s\wedge tf(t)dt\] then we can write (1) as \begin{equation}K=\left(-\Delta\right)^{-1}\end{equation}  and in this way we see that starting from $B_t$, we uniquely determine $-\Delta$ as the inverse of the integral operator $K$ defined by the covariance of $B_t$.  
\begin{figure}[!t]
\centering
\caption{fractional Gaussian bridge with $\alpha=.1$, $.5$, $.9$ from top to bottom.}
\includegraphics[width=2.5in]{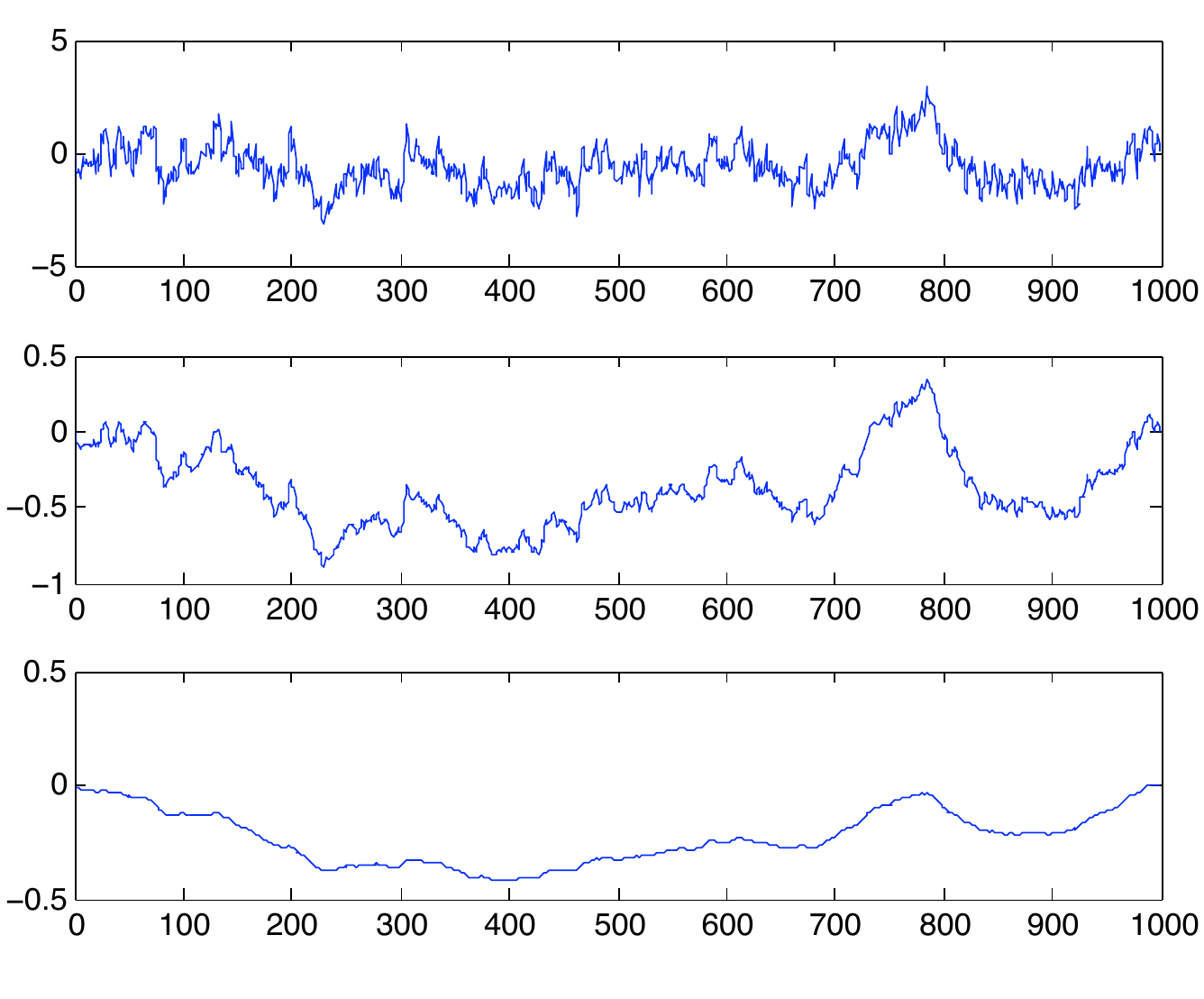}
\end{figure}
If $\{\lambda_k\}_{k=1}^{\infty}$ and $\{\phi_k\}_{k=1}^\infty $ are the eigenvalues and eigenfunctions respectively of $-\Delta$, with the above boundary conditions, then a calculation shows \[\lambda_k=\left(k-\frac12\right)^2\pi^2,\quad \phi_k(x)=\sqrt2\sin\left(k-\frac12\right)\pi x.\] The Spectral Theorem and the functional calculus associated with it then yield \[-\Delta(f)=\sum_{k=1}^\infty\lambda_k\<f,\phi_k\>\phi_k,\] and \[K(f)=\sum_{k=1}^\infty\lambda_k^{-1}\<f,\phi_k\>\phi_k,\] $\<f,g\>$ denoting the $L^2$ inner product, $\int_0^1f\bar gdx$.  If we now write \[K(f)=\int_0^1k(x,y)f(y)dy,\]\[k(x,y)=\sum_{k=1}^\infty\lambda_k^{-1}\phi_k(x)\phi_k(y),\]  it is easily seen that  the series defining $k(x,y)$ converges absolutely and uniformly on $[0,1]$ and moreover it must be that $k(x,y)=x\wedge y$.  Then if $\{\xi_k\}$ are i.i.d$.$ standard normal random variables, an application of Fubini's Theorem yields
\[\E\left[\left(\sum_{k=1}^\infty\lambda_k^{-\frac12}\xi_k\phi_k(s)\right)\left(\sum_{k=1}^\infty\lambda_k^{-\frac12}\xi_k\phi_k(t)\right)\right]=k(s,t)=s\wedge t\] and
we thus arrive at the well known Fourier series expansion of $B_t$,\begin{equation}B_t=\sum_{k=1}^\infty\lambda_k^{-\frac12}\xi_k\phi_k(t)=\sqrt2\sum_{k=1}^\infty\xi_k \frac{\sin\left(k-\frac12\right)\pi t}{\left(k-\frac12\right)\pi}.\end{equation}

Now let $W$ be the white noise (or isonormal process, cf \cite{MR1474726}) on $L^2[0,1]$ and denote its action a function $f\in L^2$ by \[W(f)\equiv\int fdW.\]  Then $\{W(\phi_k)\}$ is a set of i.i.d$.$ standard normal random variables, denoted again by $\{\xi_k\}$, and if \begin{align}\notag k^\frac12(x,y)&=\sum_{k=1}^\infty\lambda_k^{-\frac12}\phi_k(x)\phi_k(y)\\\notag&=\sqrt2\sum_{k=1}^\infty \frac{\sin\left(\left(k-\frac12\right)\pi x\right)\sin\left(\left(k-\frac12\right)\pi y\right)}{\left(k-\frac12\right)\pi}\end{align} then \[\left(-\Delta\right)^{-\frac12}f(x)=\int_0^1k^\frac12(x,y)f(y)dy\] and \begin{align}\notag\left(-\Delta\right)^{-\frac12}\left(W\right)&=\int k^\frac12(x,y)dW(y)\\\notag&=\sum_{k=1}^\infty\lambda_k^{-\frac12}\int\phi_k(y)dW\phi_k(x)\\\notag&=\sum_{k=1}^\infty\xi_k \frac{\sin\left(k-\frac12\right)\pi t}{\left(k-\frac12\right)\pi}.\end{align}  Thus we can write $B_t=\left(-\Delta\right)^{-\frac12}\left(W\right)$ and arrive at the stochastic steady-state equation \begin{equation}\left(-\Delta\right)^\frac12B_t=W.\end{equation}  Notice that working backwards we can \textit{define} $B_t$ to be the unique Gaussian process satisfying (4).  This is similar, but not the same as, the classical equation \[\frac d{dt}B_t=W.\]  To see that (4) is in fact different, note that $\left(-\Delta\right)^{-\frac12}$ is self adjoint on $L^2[0,1]$, while $d/dt$ is not (integrate by parts).  

Now suppose we replace the mixed Laplacian above by the Dirichlet Laplacian in (4), acting on functions such that $f(0)=f(1)=0$.  The Dirichlet Laplacian on $[0,1]$ has eigenvalues and eigenfunctions given by \[\lambda_k=(k\pi)^2\] and \[\sqrt2\sin(k\pi x)\] respectively.  Then we have as above a process $X_t$ defined by \[X_t=\sum_{k=1}^\infty\xi_k\frac{\sin(k\pi t)}{k\pi}.\]  If we now let $\alpha\in(0,1)$ (the case above being $\alpha=1/2$) we can extend (4) to \begin{equation}\left(-\Delta\right)^{\left(\frac 14+\frac\alpha2\right)}X_t=W\end{equation} and write \begin{equation}X_t=\sum_{k=1}^\infty\xi_k\frac{\sin(k\pi t)}{(k\pi)^{\left(\frac12+\alpha\right)}},\end{equation} which is the \textit{homogeneous Riesz field} constructed in \cite{FBFOM}.  This is a self-similar Gaussian process with almost surely Holder continuous sample paths and such that $X_0=X_1=0$ almost surely, i.e., it is a Gaussian bridge. 

Because the series converges in $L^2$ with probability $1$, we can simulate these processes by taking partial sums (see fig. 1).  Notice that as $\alpha$ increases the sample paths become more regular.  The reason can be seen from (6):  Lower alpha emphasizes the larger eigenvalues and thus the higher frequency wave functions show more of a contribution.  Similarly a larger $\alpha$ suppresses their contribution and the lower frequency wave functions dominate, leading to less overall oscillation and a smoother sample path.  We will see this same behavior in general below.
\begin{figure}[!t]
\centering
\subfloat[$\alpha=.1$]{\includegraphics[width=2.5in]{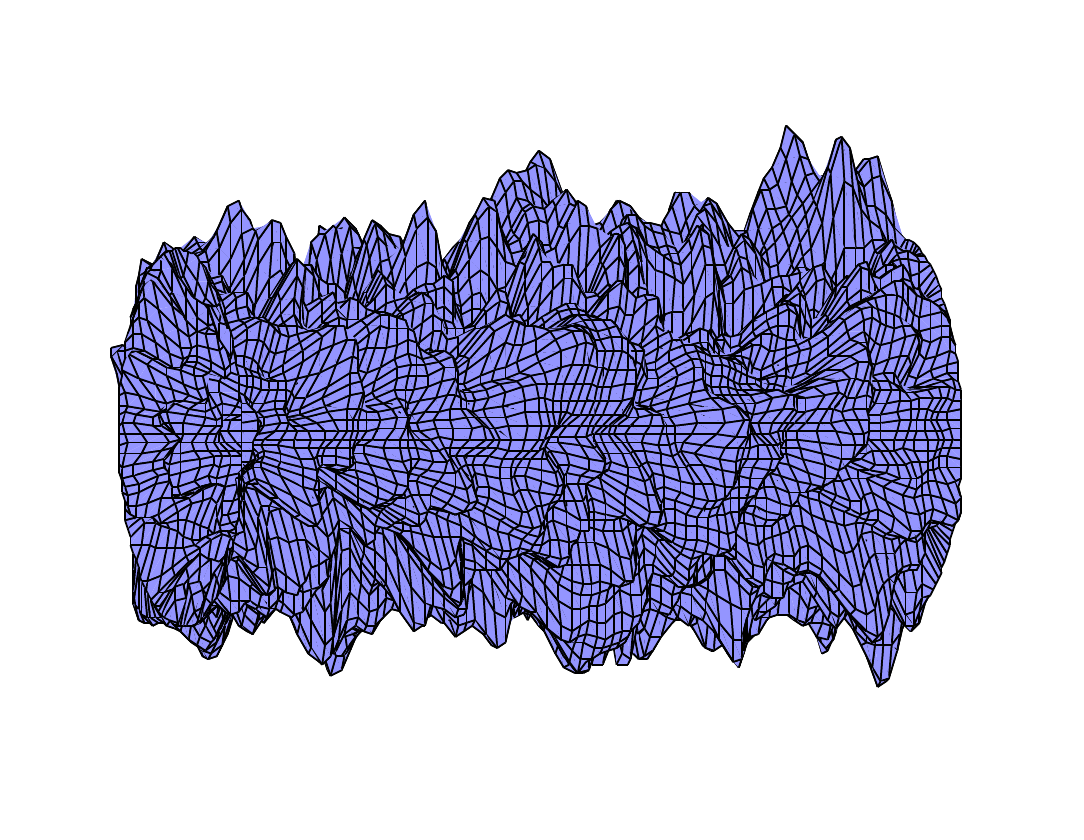}%
\label{fig_first_case}}\\
\subfloat[$\alpha=.5$]{\includegraphics[width=2.5in]{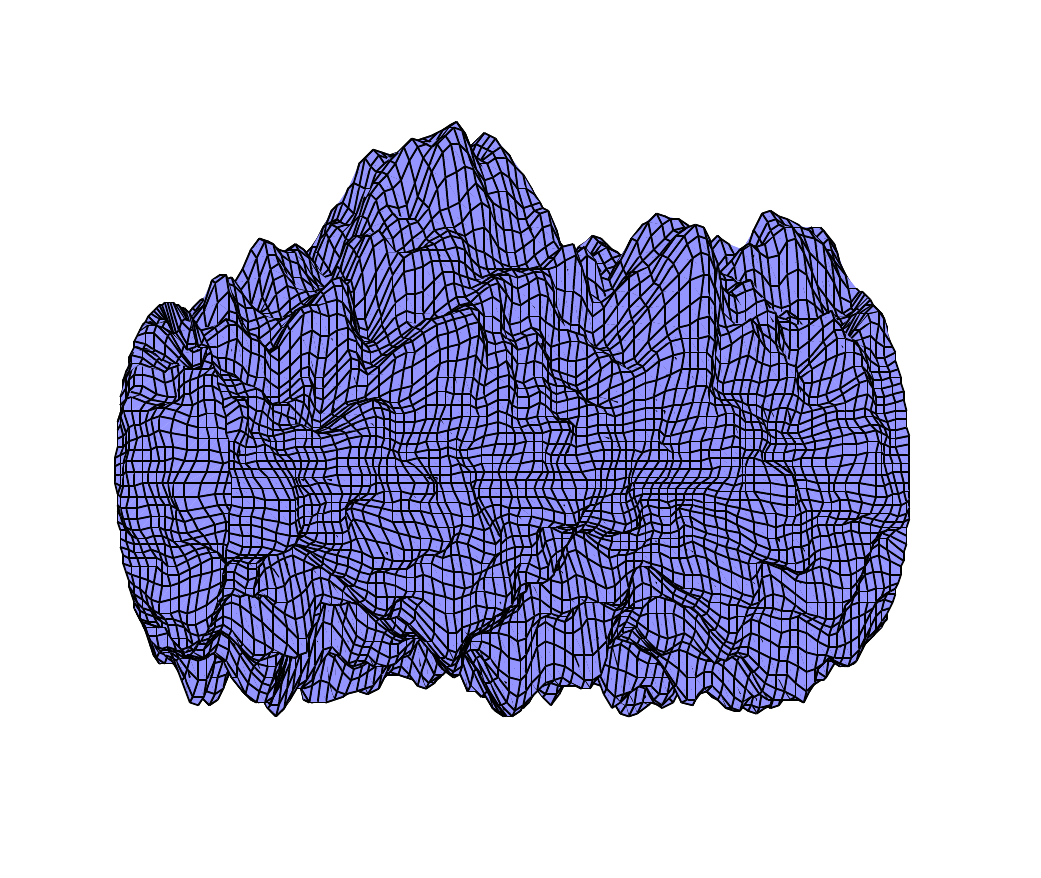}%
\label{fig_second_case}}\\
\subfloat[$\alpha=.9$]{\includegraphics[width=2.5in]{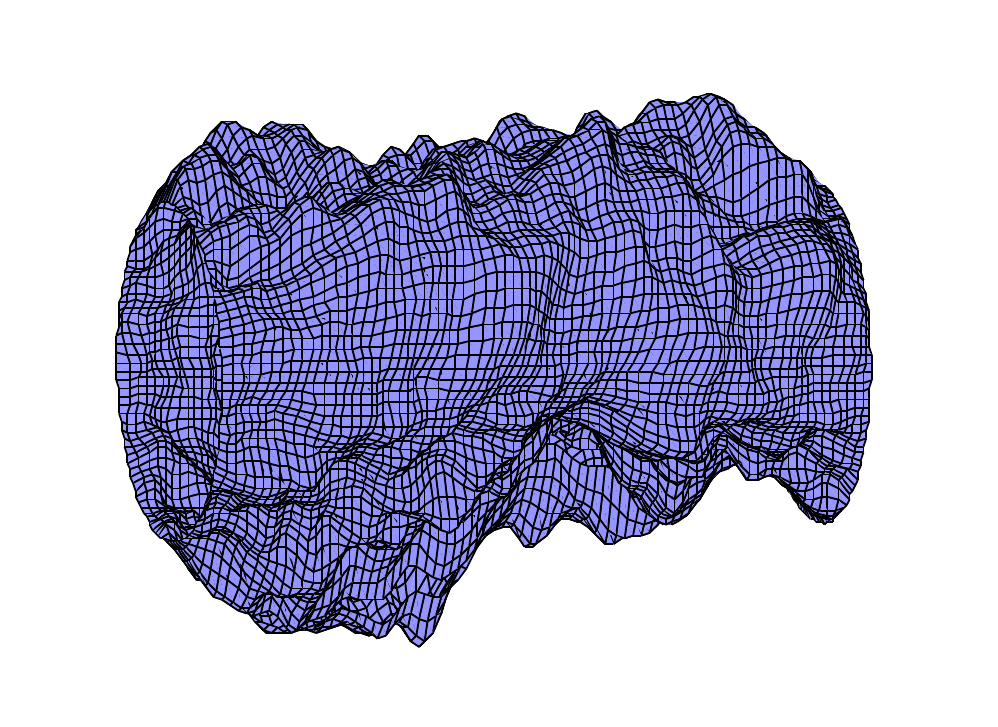}%
\label{fig_second_case}}
\caption{Cylinders with Dirichlet boundary conditions for $\alpha=.1, .5, .9$}
\label{fig_sim}
\end{figure}

\subsection{Extension to Surfaces with Boundary}
Suppose now $D$ is a compact connected surface in $\R^3$ with smooth boundary.  Then there is a canonical differential operator on $C^\infty(D)$ corresponding to the Euclidean Laplace operator above, the Laplace-Beltrami operator, which we refer to also as simple the Laplacian of $D$ (see e.g. \cite{MR768584}).  As above we let $\lambda_k$ and $\phi_k$ be the eigenvalues and eigenfunctions of the Dirichlet Laplacian of $D$, $-\Delta$.  Given a white noise $W$ on $D$ we start from \[\left(-\Delta\right)^{\left(\frac 12+\frac\alpha2\right)}R_x=W\] to obtain \begin{equation}R_x=\sum_{k=1}^\infty\lambda_k^{-\left(\frac12+\frac\alpha2\right)}\xi_k\phi_k(x),\end{equation} where the change in the exponent from $1$ to $2$ ensures sample path regularity (see \cite{FBFOM}).  In this way we obtain a random field on $D$ that is self-similar, almost-surely H\"older continuous of order $\gamma$ for any $\gamma<\alpha$ and almost surely not H\"older continuous for $\gamma\geq\alpha$, and has stationary increments.  A few words are in order as to what we mean by self-similar:  Because we are considering surfaces embedded in $\R^3$, that is, surfaces with Riemannian metrics induced by the Euclidean metric on $\R^3$, the usual definitions carry over, i.e., $R_{cx}\stackrel{d}=c^\alpha R_x$.  For more general considerations of self-similarity on arbitrary manifolds see \cite{FBFOM}.

Now we have a generalization of the bridges above that possesses the properties we would like, however for a general surface $D$ we no longer have simple analytical expressions for the functions $\phi_k$.  One way around this is to discretize $\Delta$ and construct the analogous discretized version of (7).  This requires some care, as there are a number of discretizations available for Laplace-Beltrami operators, some of which behave very differently (e.g. \cite{freelunch}).

Given a meshed surface, most discrete Laplacians are weighted graph Laplacians, \[Lf(x)=\sum_{x\sim y} w_{x,y}(f(x)-f(y))\] for some edge weights $w_{x,y}$ and where $x\sim y$ means $x$ and $y$ are adjacent in the mesh (i.e$.$ they share an edge).  Supposing we have a fairly uniform mesh, let us set $w_{x,y}=\|x-y\|_{\R^3}^{-2}$.  Denote this discrete Laplacian by $L$, and let its eigenvectors and eigenvalues be $\{\bar\phi_k\}$ and $\{\bar \lambda_k\}$ respectively.  Note that $\bar\phi_k\in \R^N$ where the dimension of $L$ ($L$ is a matrix) is $N\times N$, and $\phi_k$ is a real valued function on our mesh.  We then let \begin{equation}\bar R_x=\sum_{k=1}^{N_0}\bar\lambda_k^{-\left(\frac12+\frac\alpha2\right)}\xi_k\bar\phi_k(x),\end{equation} where we choose some $N_0\leq N$ and $\xi_k$ are again independent standard normals.

We would like to be able to take our  mesh fine enough and $N_0$ large enough so that $\bar R_x$ provides a good approximation of the full Riesz field $R_x$.  How good this approximation is, and in what sense, will depend completely on how well $L$ approximates $\Delta$.  For general surfaces, discrete Laplacians and their convergence to the continuum Laplacian is a very active area of current research, and we defer a more full discussion for the moment.
\begin{figure}[!t]
\centering
\subfloat{\includegraphics[width=2.5in]{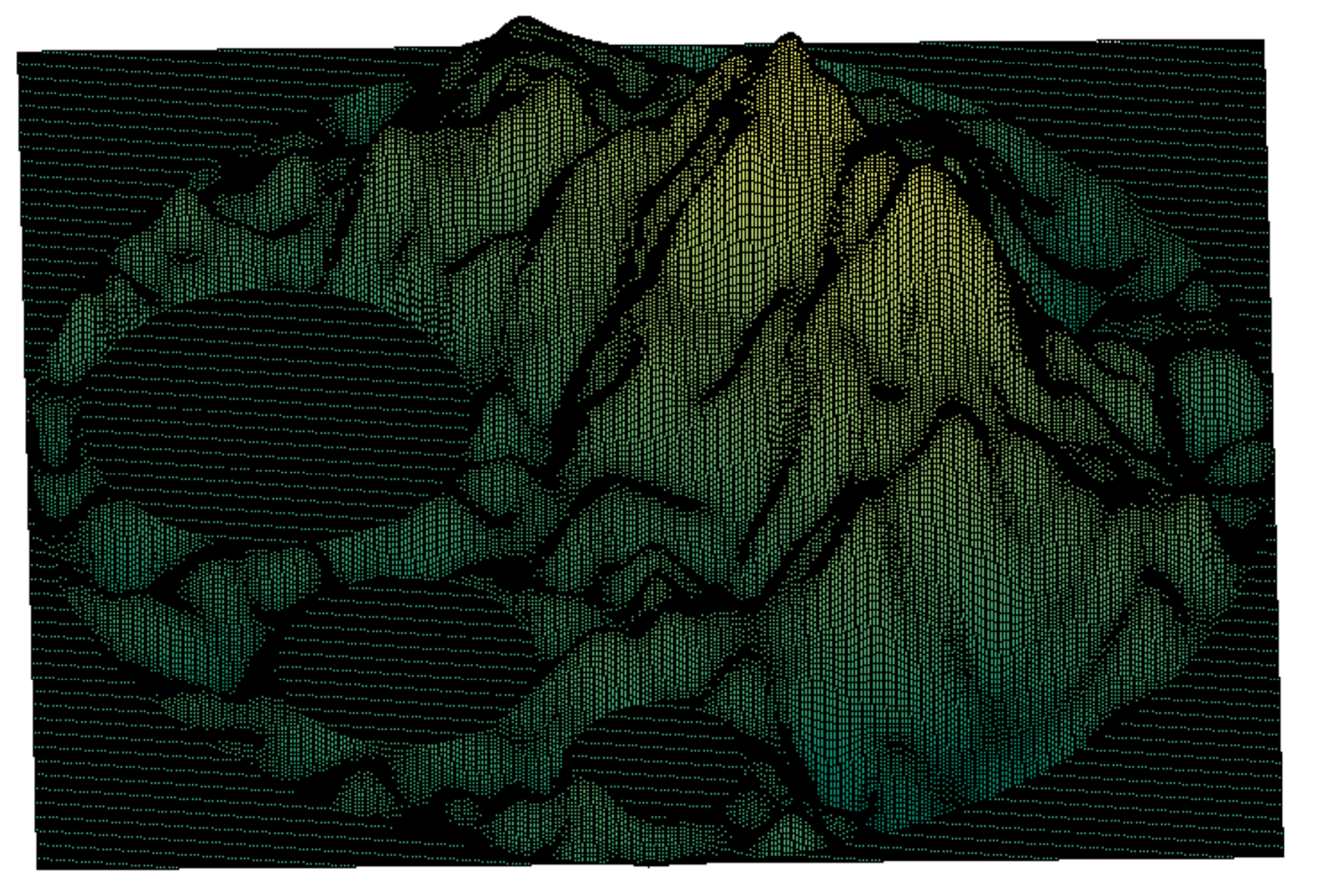}%
\label{fig_first_case}}\\
\subfloat{\includegraphics[width=2.5in]{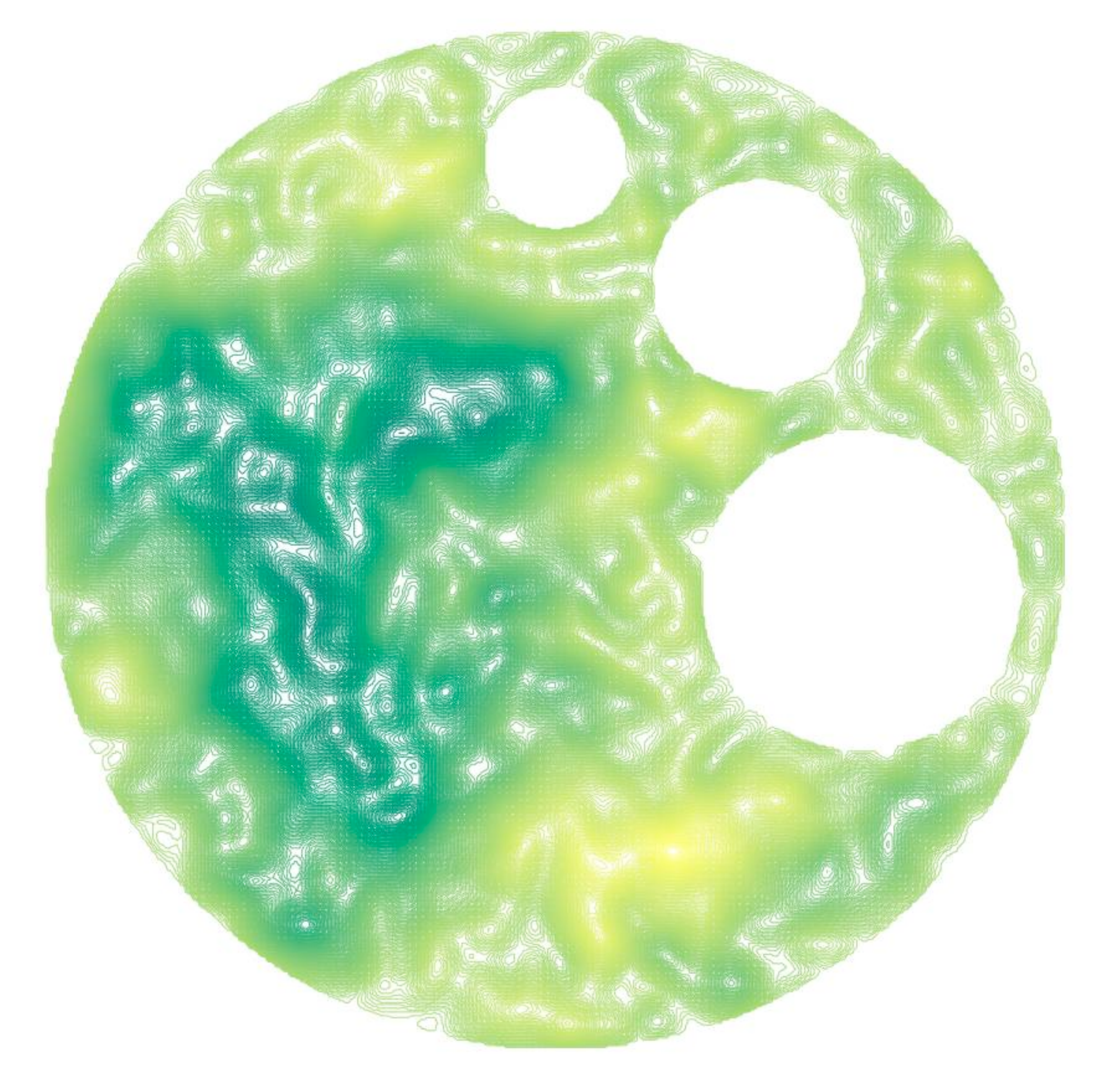}%
\label{fig_second_case}}\\
\caption{Disk with 3 smaller disks removed, $\alpha=.9$}
\label{fig_sim}
\end{figure}

Some examples of planar domains are plotted in Figs. 3, 5, and 6.  Note that with probability $1$ the field is zero on the boundary.  As an example of what is possible in three dimensions, consider the cylinder $[0,C]\times \mathbb S^1$ (Fig. 2).  If we impose zero Dirichlet boundary conditions on the two boundary circles, we get analogues of the bridges in Fig. 1.  Notice that again, $\alpha$ controls the sample path (or surface) regularity and the fields are almost surely equal to zero on both boundary circles, i.e., the height of the field values plotted along normals is zero.

\subsection{Surfaces Without Boundary}
For surfaces without boundary there are no obvious boundary conditions, so we need to interpret our equations differently.  The basic issue is that on a compact manifold, e.g$.$, the sphere $\mathbb S^2$, the lowest eigenvalue of the Laplacian is zero and $(-\Delta)^{-(\frac12+\frac\alpha2)}$ is not defined.  We can instead consider the series \begin{equation}R^\alpha_x\stackrel {d}=\sum_{k=1}^\infty (\lambda_k)^{-\left(\frac d4+\frac\alpha2\right)}(\phi_k(x)-\phi_k(o))\xi_k\end{equation} where $o$ is some fixed point serving as an ``origin."  This series does converge and again defines a self-similar field with stationary increments and almost sure H\"older continuity as above, called the \textit{Riesz Field} in \cite{FBFOM}.  Figure 4 shows spheres with increasing alpha, using the same discrete Laplacian $L$ as before.
\begin{figure}[!t]
\centering
\subfloat[$\alpha=.1$]{\includegraphics[width=2.5in]{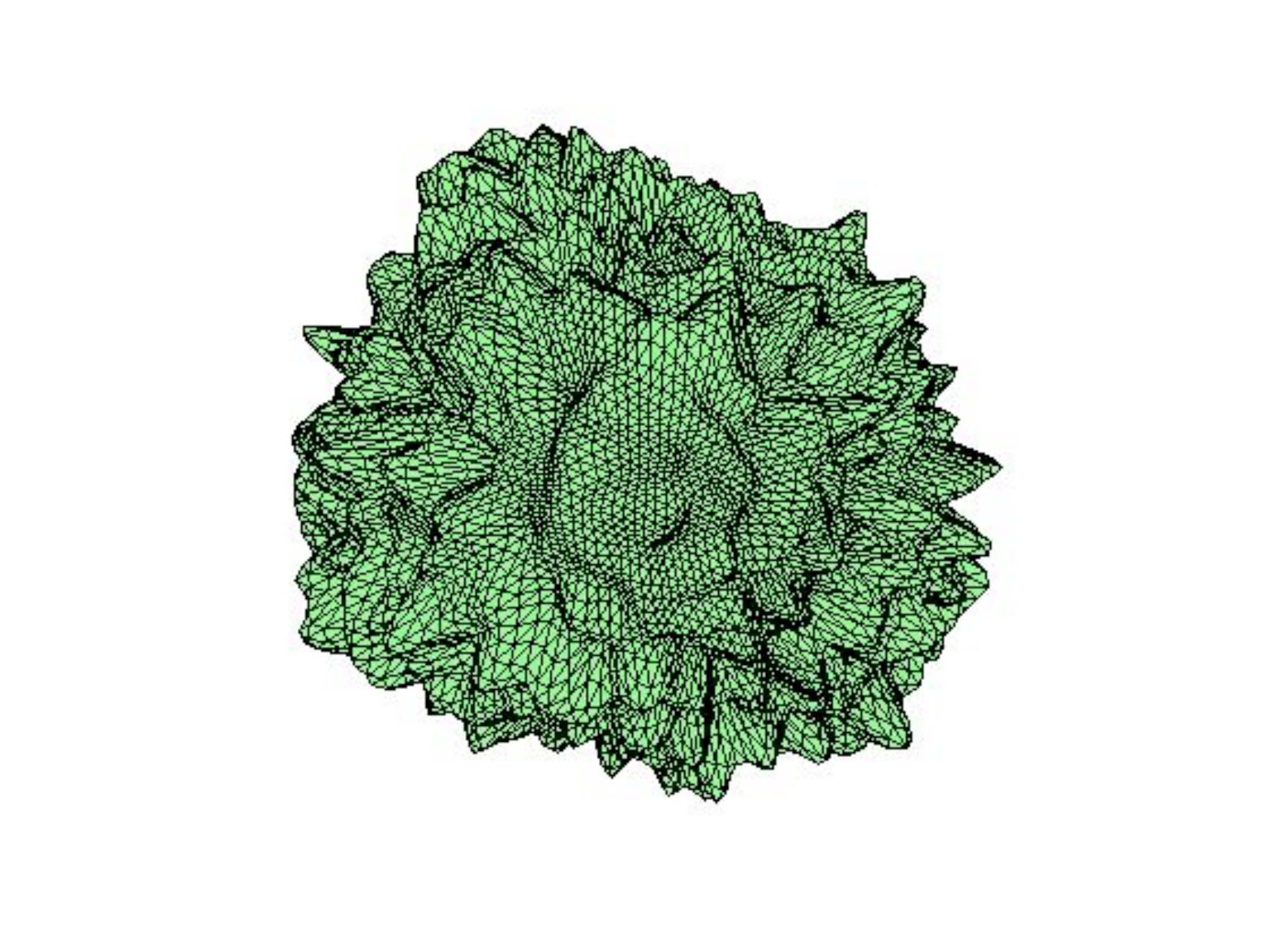}%
\label{fig_first_case}}\\
\subfloat[$\alpha=.5$]{\includegraphics[width=2.5in]{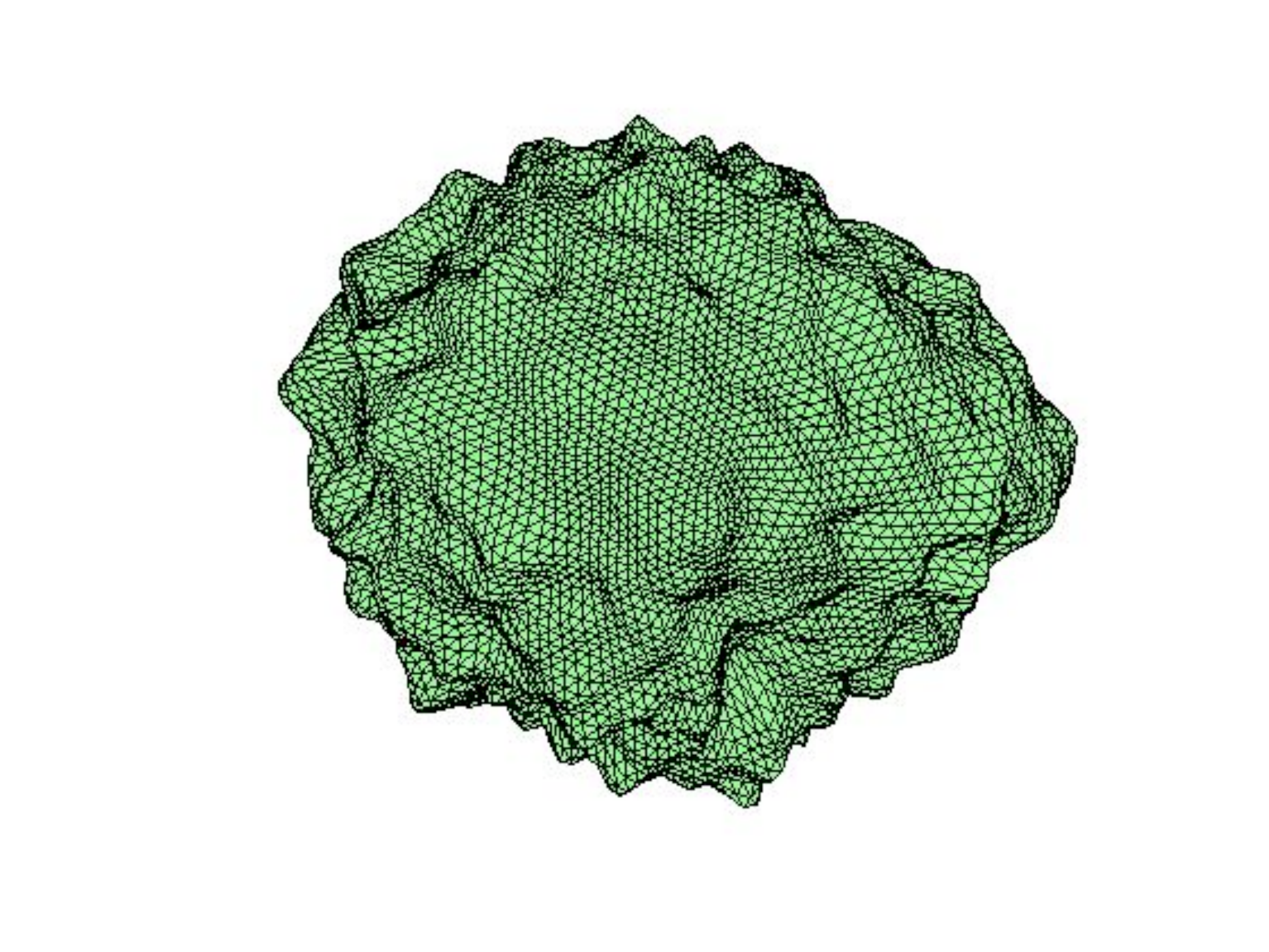}%
\label{fig_second_case}}\\
\subfloat[$\alpha=.9$]{\includegraphics[width=2.5in]{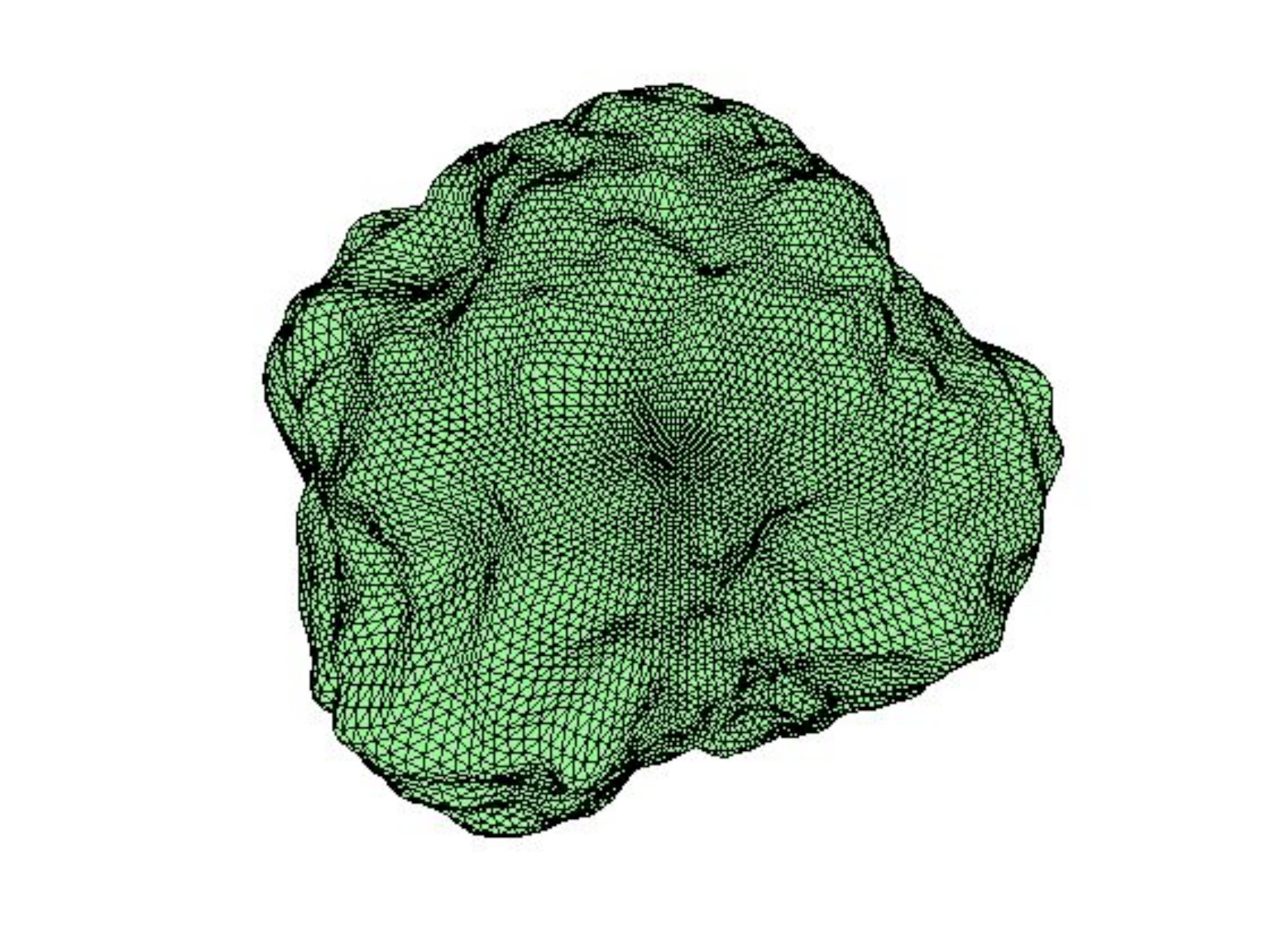}%
\label{fig_second_case}}
\caption{Spheres with $\alpha=.1, .5, .9$}
\label{fig_sim}
\end{figure}

\section{Implementation}
\subsection{Convergence}
For uniform or nearly uniform meshes, the above discretization produces good results.  In general however, placing a uniform mesh on a surface is non-trivial (depending on the curvature and boundary), and in this case the above discrete Laplacian may be a poor approximation to the continuous Laplacian.  In the case of highly curved surfaces, care must be taken in choosing an appropriate discretization.

Suppose we have a meshed surface and have chosen a discrete Laplacian, $L$.  A simple condition for the almost sure $L^2$ approximation $\|R-\bar R\|_{L^2}<\epsilon$ is as follows:  Suppose that for any integer $N>0$, for all $n\leq N$ we have \[\lim_{\mbox{mesh}\to0}\|\bar\phi_n-\phi_n\|_{L^2}=0\] and \[|\bar\lambda_n-\lambda_n|\to0,\]  where we can embed $\bar\phi_k$ in $L^2(D, dV)$, $dV$ denoting surface measure on $D$.  Then for any $\epsilon>0$ we can choose $N_\epsilon$ such that with probability 1 \begin{align}\notag\lim_{\mbox{mesh}\to0}&\left\|\sum_{n=1}^{N\epsilon}\bar\lambda_n^{-(\frac d4+\frac\alpha2)}\bar\phi_n(x)\xi_n-\sum_{n=1}^\infty\lambda_n^{-(\frac d4+\frac\alpha2)}\phi_n(x)\xi_n\right\|_{L^2}\\\notag&<\epsilon.\end{align}  This is a simple consequence of the fact that \[\sum_{n=1}^\infty\lambda_n^{-(\frac d4+\frac\alpha2)}\phi_n(x)\xi_n\] converges in $L^2$ almost surely, for then we just bound the tail and the assumed convergence of the lower eigenfunctions and eigenvalues yields the claim.  

Thus if one has a meshed surface, the Laplacian chosen should satisfy the above spectral convergence condition.  Such conditions are known to hold for certain Laplacians, but not all \cite{MR2768624, ISI:000267277500022}.  Of course, this is an active area of research and so we expect further results on spectral approximation of the Laplace-Beltrami operator to become applicable to the spectral synthesis we have used here.
\subsection{Efficiency and Self-Similarity}
There are two parameters which determine the computation time in simulating $R_x$ for a given surface:  The number $n$ of grid points in the mesh (i.e. how fine the mesh is) and the number $N$ of eigenvectors appearing in \[\bar R_x=\sum_1^N\left(\bar\lambda_k\right)^{-\left(\frac12+\frac\alpha2\right)}\xi_k\bar\phi_k(x).\]  The dimension of the discrete Laplacian will be $n\times n$ and in general $N$ should be taken much smaller than $n^2$, for Weyl's asymptotic formula for the eigenvalues of $\Delta$ (e.g$.$ \cite{MR768584}) tells us that $\lambda_k=O(k)$ for a two dimensional surface as we are dealing with here.  Thus \[\lambda_k^{-\left(\frac12+\frac\alpha2\right)}=O\left(k^{-\left(\frac12+\frac\alpha2\right)}\right)\] and so if our discrete Laplacian $L$ satisfies the spectral convergence above, we see that the contribution, in mean square, of the $N$th term is of the order $N^{-\left(\frac12+\frac\alpha2\right)},$ which may rapidly become negligible depending on the fineness of the mesh.

On the other hand, for finer meshes, the computation of even the first few hundred eigenvectors and eigenvalues of the $n\times n$ matrix $L$ can become long.  However, once computed, this spectral data can be stored for repeated use.  All that is needed is the random numbers $\{\xi_k\}$, which are not expensive.  Moreover, the self-similarity of the Riesz fields allows one to compute the spectra of a surface of any size, and simply rescale the resulting field without computing the spectra for a rescaled surface.  For example, once one has simulated the Riesz field on a sphere of radius $1$, one can simulate it on a sphere of radius $c$ simply by plotting $c^\alpha \bar R_x$ for the appropriate value of $\alpha$.  In this way, for very detailed images, the computational time required may be somewhat high, but it is a one time cost in the above sense.
\begin{figure}[]
\centering
\caption{Fractal surface on the unit disk with Dirichlet boundary conditions ($\alpha=.7$) with negative values set to zero.}
\includegraphics[width=2.5in]{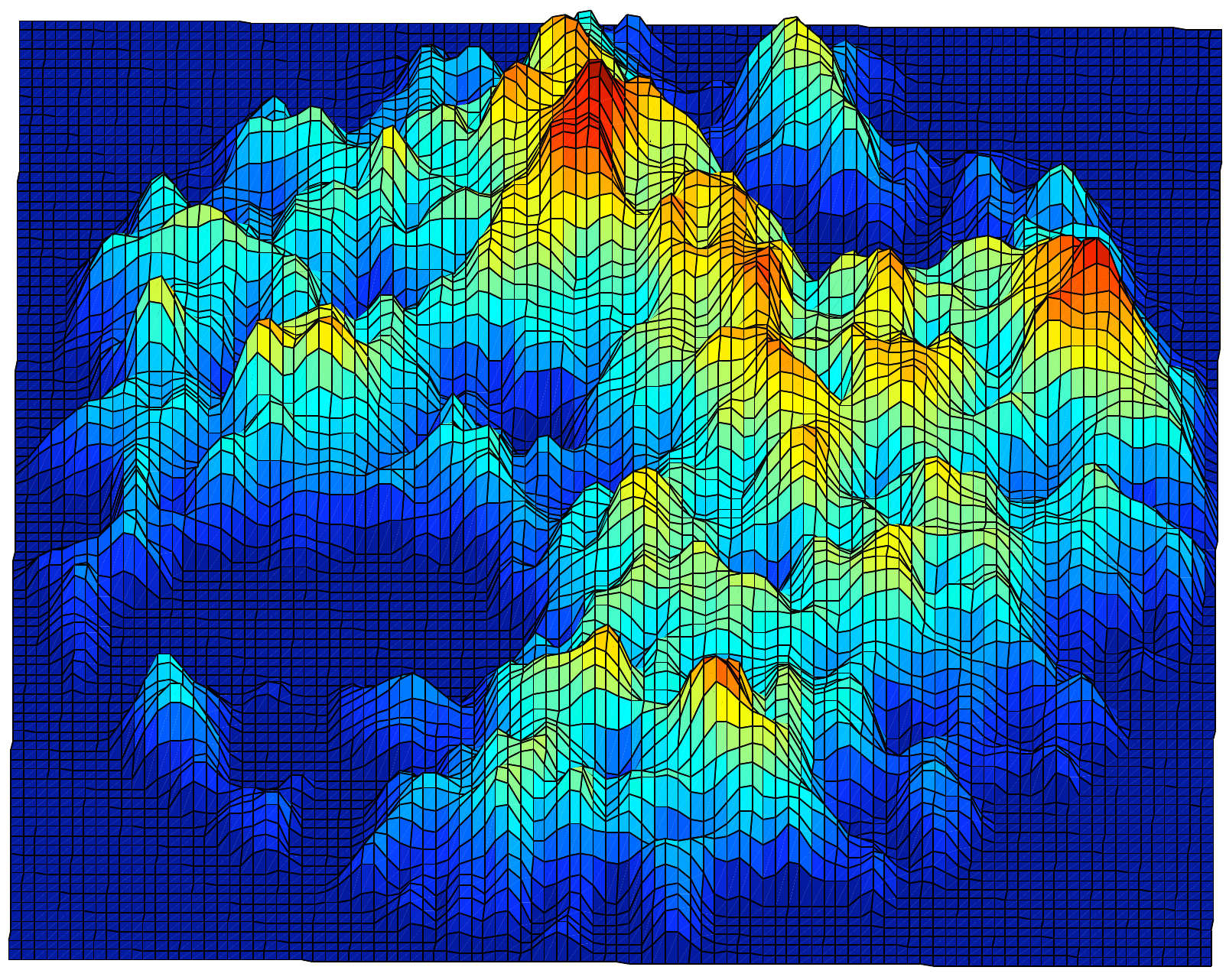}
\end{figure}
\begin{figure}
\centering
\subfloat[]{\includegraphics[width=2.5in]{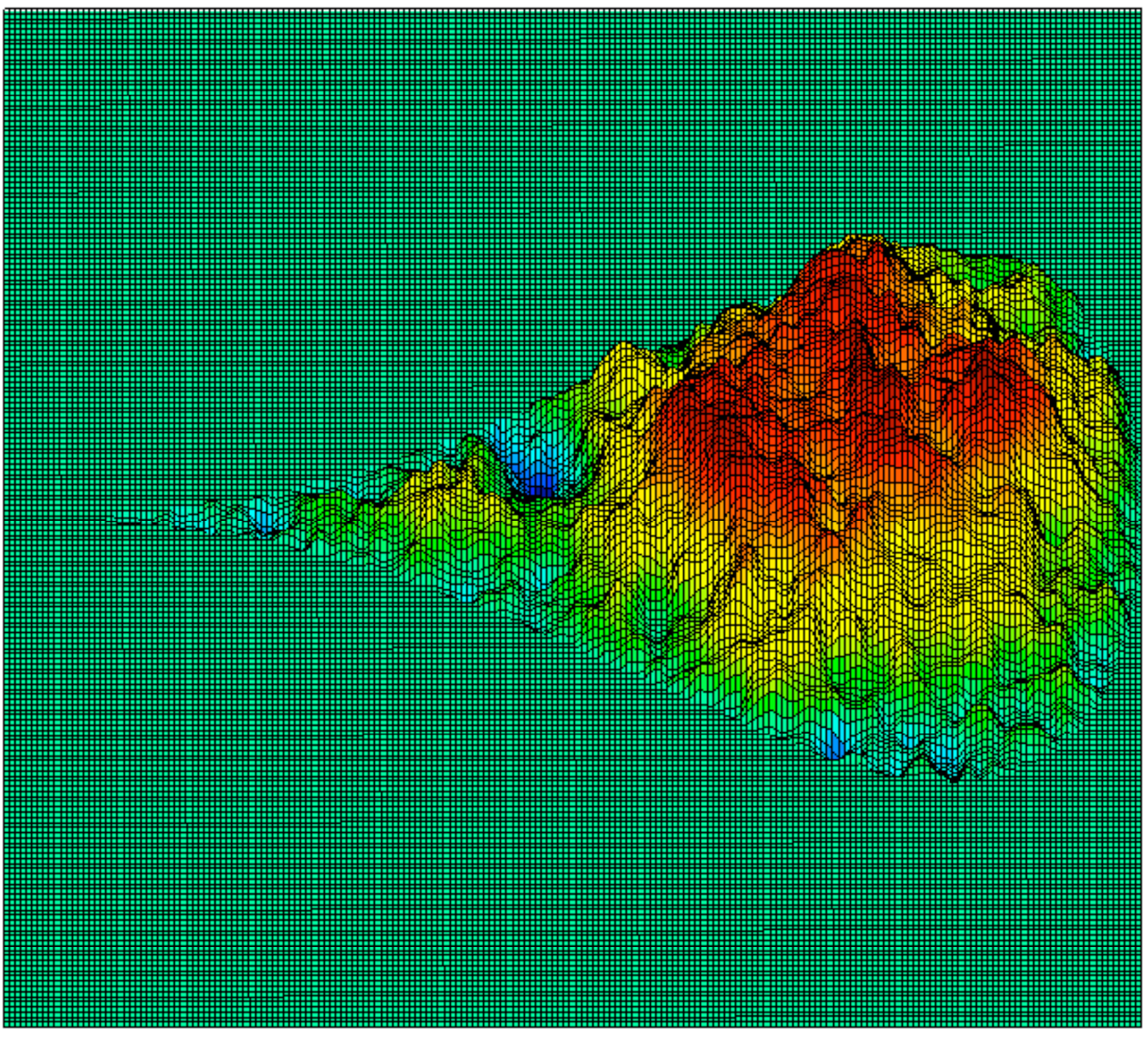}%
\label{fig_first_case}}\\
\subfloat[]{\includegraphics[width=2.5in]{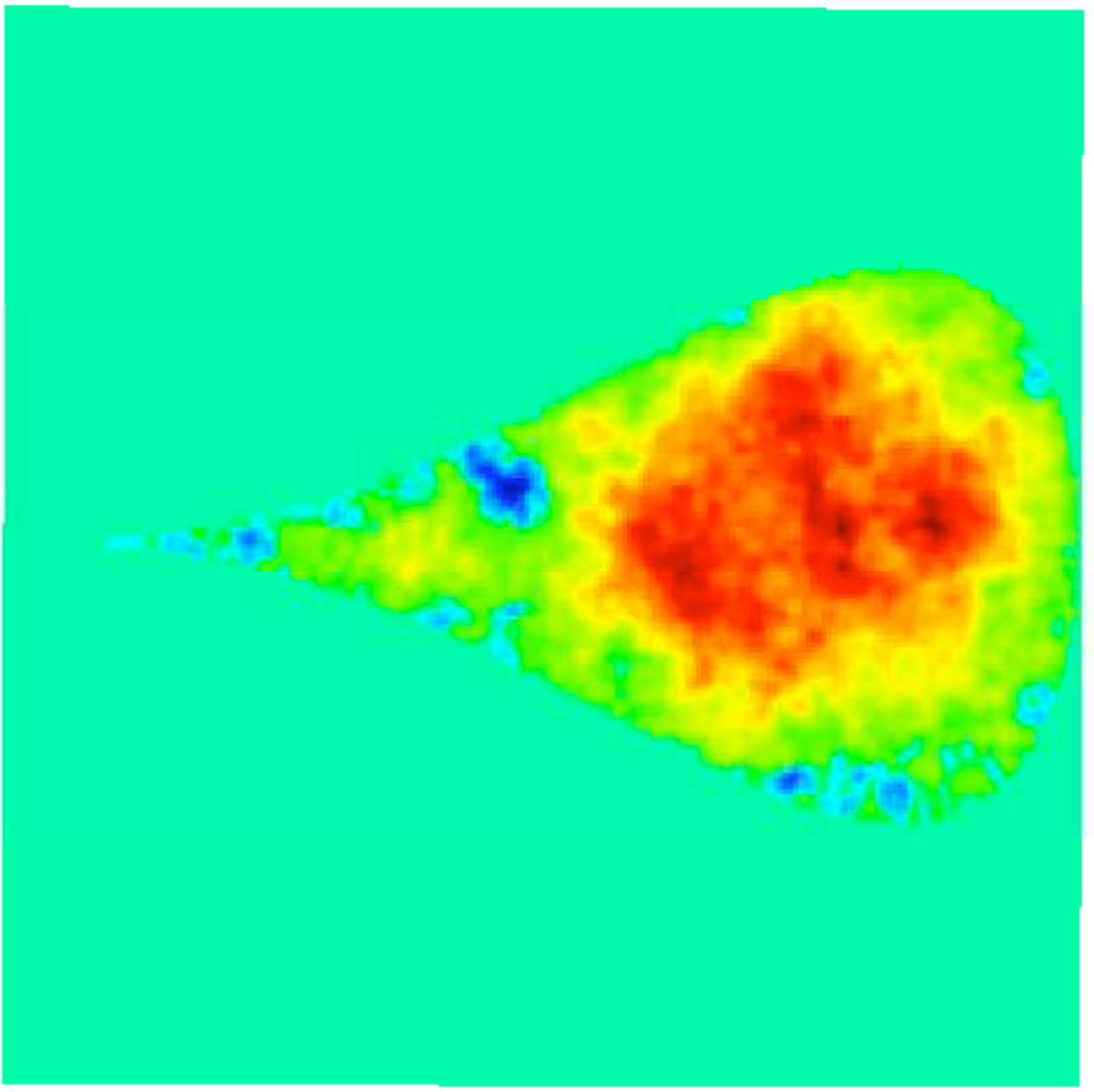}%
\label{fig_second_case}}
\caption{Fractal surface over a teardrop shaped region with $\alpha=.7$.}
\label{fig_sim}
\end{figure}
\section{Possible Extensions}
We have restricted our attention to embedded surfaces in $\R^3$, however the theory in \cite{FBFOM} applies to non-embedded manifolds, e.g., hyperbolic space or flat tori, as well as $n$-dimensional surfaces in $\R^{n+1}$.  In this case the approximation takes the form \[\bar R_x=\sum_1^N\left(\bar\lambda_k\right)^{-\left(\frac n4+\frac\alpha2\right)}\xi_k\bar\phi_k(x).\]  Of course working with a surface whose metric is not induced by the ambient Euclidean metric increases the difficulty in computing accurate discretizations of the Laplacian.  Also, we have here chosen $\alpha$ to be fixed, but one could also allow $\alpha$ to vary over the surface and thus obtain multifractal fields.  We expect, similar to what is known on $\R^n$, that such fields with varying fractal index would still be continuous.



\section*{Acknowledgment}  The first author thanks his advisor, Harold Parks, for his time and encouragement.  Figure 4 was produced using the Matlab Toolboxes ``Toolbox Graph'' and ``Toolbox Wavelets on Meshes,'' written by Gabriel Pyre.

\ifCLASSOPTIONcaptionsoff
  \newpage
\fi



\bibliographystyle{IEEEtran}
\bibliography{FBS(IEEE).bib}
%


\IEEEbiographynophoto{Oregon State University\\gelbaumz@math.oregonstate.edu\\titusm@math.oregonstate.edu}
%







\end{document}